\documentclass[conference]{IEEEtran}

\usepackage{cite}

\ifCLASSINFOpdf
  \usepackage[pdftex]{graphicx}
  \graphicspath{{pdf/}{jpeg/}{media/}}
  \DeclareGraphicsExtensions{.pdf,.jpeg,.png}
\else
\fi

\usepackage{url}
\usepackage{textcomp}
\usepackage{booktabs}
\usepackage{tabularx}
\usepackage[utf8]{inputenc}
\usepackage[T1]{fontenc}
\usepackage{float}
\usepackage{footnote}
\usepackage{subfigure}
\usepackage{color}
\usepackage{listings}

\usepackage[frozencache=true,cachedir=.]{minted}
\usepackage{fancyhdr}
\fancypagestyle{firstpage}{%
  \rhead{\bfseries \textcolor{red}{Presented at Crypto Valley Conference on Blockchain Technology (CVCBT 2018), 20-22 June 2018\\Published version may differ}}
}

\begin{document}

\title{IDMoB: IoT Data Marketplace on Blockchain}

\author{\IEEEauthorblockN{Kaz{\i}m R{\i}fat \"{O}zy{\i}lmaz} 
\IEEEauthorblockA{Computer Engineering Department\\
Bogazici University\\
Istanbul, Turkey\\
kazim@monolytic.com}
\and
\IEEEauthorblockN{Mehmet Do\u{g}an}
\IEEEauthorblockA{Computer Engineering Department\\
Bogazici University\\
Istanbul, Turkey\\
mehmet.dogan1@boun.edu.tr}
\and
\IEEEauthorblockN{Arda Yurdakul} 
\IEEEauthorblockA{Computer Engineering Department\\
Bogazici University\\
Istanbul, Turkey\\
yurdakul@boun.edu.tr}}

\maketitle

\thispagestyle{firstpage}

\begin{abstract}
Today, Internet of Things (IoT) devices are the powerhouse of data generation with their ever-increasing numbers and widespread penetration. Similarly, artificial intelligence (AI) and machine learning (ML) solutions are getting integrated to all kinds of services, making products significantly more "smarter". The centerpiece of these  technologies is "data". IoT device vendors should be able keep up with the increased throughput and come up with new business models. On the other hand, AI/ML solutions will produce better results if training data is diverse and plentiful.

In this paper, we propose a blockchain-based, decentralized and trustless data marketplace where IoT device vendors and AI/ML solution providers may interact and collaborate. By facilitating a transparent data exchange platform, access to consented data will be democratized and the variety of services targeting end-users will increase. Proposed data marketplace is implemented as a smart contract on Ethereum blockchain and Swarm is used as the distributed storage platform.
\end{abstract}

\IEEEpeerreviewmaketitle

\section{Introduction}
A new age of always listening, monitoring and communicating IoT devices are at our doorstep. Quoting IBM: "90 percent of the data in the world today has been created in the last two years alone – and with new devices, sensors and technologies emerging, the data growth rate will likely accelerate even more"~\cite{ibm2017marketing}. Increased amount of data enforces companies to create and maintain large scale infrastructure projects in the cloud. Unfortunately, every company, competent or not, tackles these problems in their own way. In the meantime, almost every company is building Artificial Intelligence (AI) and Machine Learning (ML) solutions. Using both publicly available and privately collected data, companies aim to provide customized user experiences targeting each individual differently. Vast amount of consented data is still not tapped and currently there is no platform to search for it.

\begin{figure}[t]
  \centering
  \includegraphics[scale=0.4]{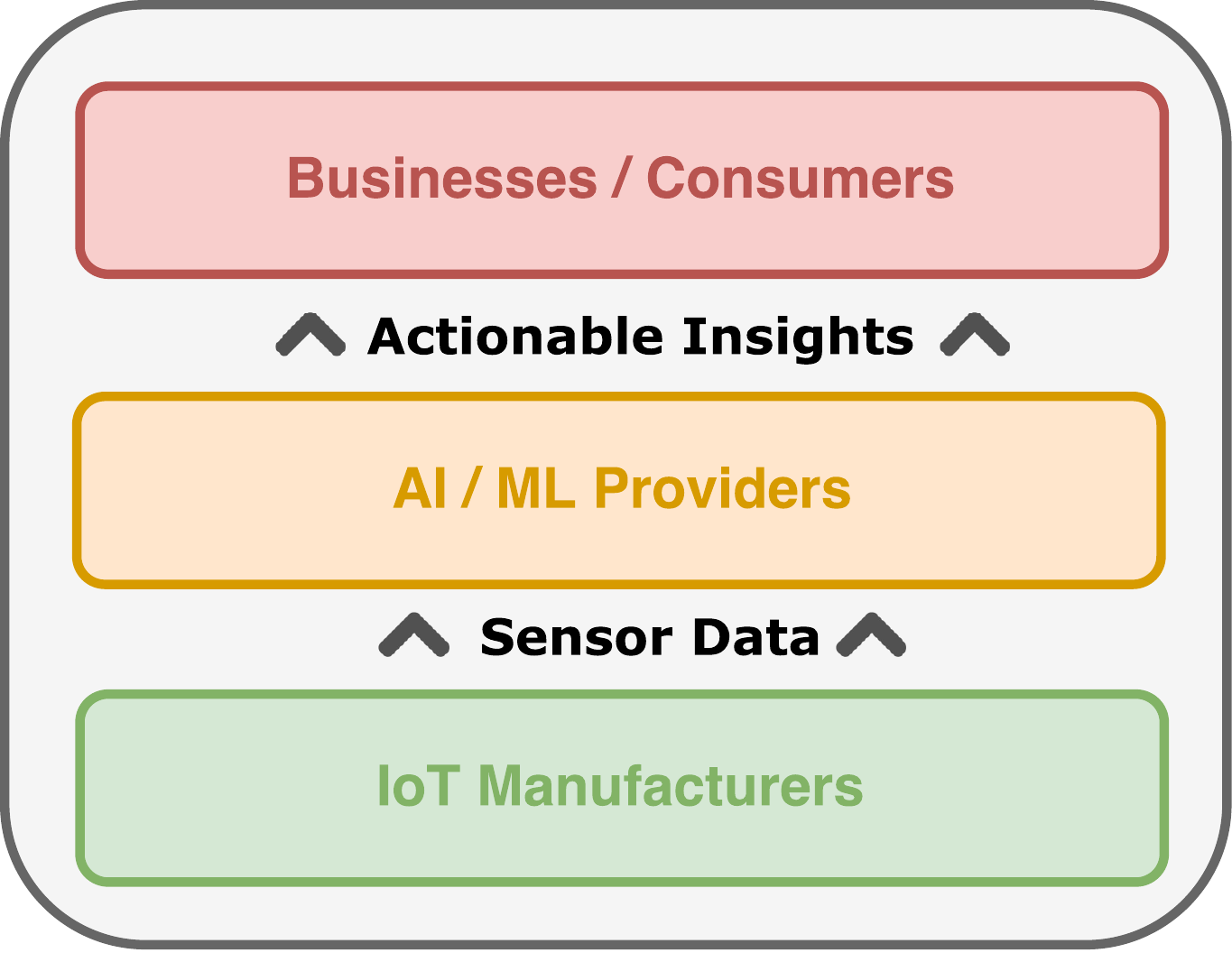}
  \caption{Multi-party, multi-layer IoT solution}
  \label{fig:stack}
\end{figure}

Today, IoT manufacturers use cloud-based solutions to implement their data storage and business intelligence/dashboard services. However implementing predictive, prescriptive and adaptive solutions necessary for future businesses requires an additional data processing step that extracts actionable triggers from all the collected data~\cite{deloitteiot}~\cite{PwCAIIoT}. These next generation of services, which is pushed by recent technology trends like Industry 4.0 and Smart Agriculture, force IoT manufacturers to develop a new skill set that they currently do not possess. On the other side, there are plenty of AI and ML startups trying to create insight using only tiny scraps of data. The solution is straightforward: having a trustable, neutral platform that data producers (IoT manufacturers) and data consumers (AI/ML providers) can seamlessly trade. We propose blockchain to facilitate such a trustless and secure digital trading platform.

Once the platform is in place, a complete “business intelligence solution” can be created just like designing a layered software stack. However, it will differ from existing solutions as follows: our solution will be consisting of multiple stakeholders (data providers and processors) that are connected to each other in a certain way using the blockchain infrastructure, to create actionable insights, i.e. information that can be acted upon, for consumers (Figure~\ref{fig:stack}). One of the many benefits of using a blockchain-based solution is that it almost always comes with a cryptocurrency attached, therefore it is very easy to make economic incentives work. Consequently, as observed in current blockchain/cryptocurrency realm (SegWit, block size debate etc.), governance is yet another big aspect that needs to be addressed in order to establish a living and working marketplace. Eliminating IoT manufacturers that provide bad data or ranking good data sets higher should be built while designing the infrastructure.

\begin{figure*}[h]
  \centering
  \includegraphics[scale=0.45]{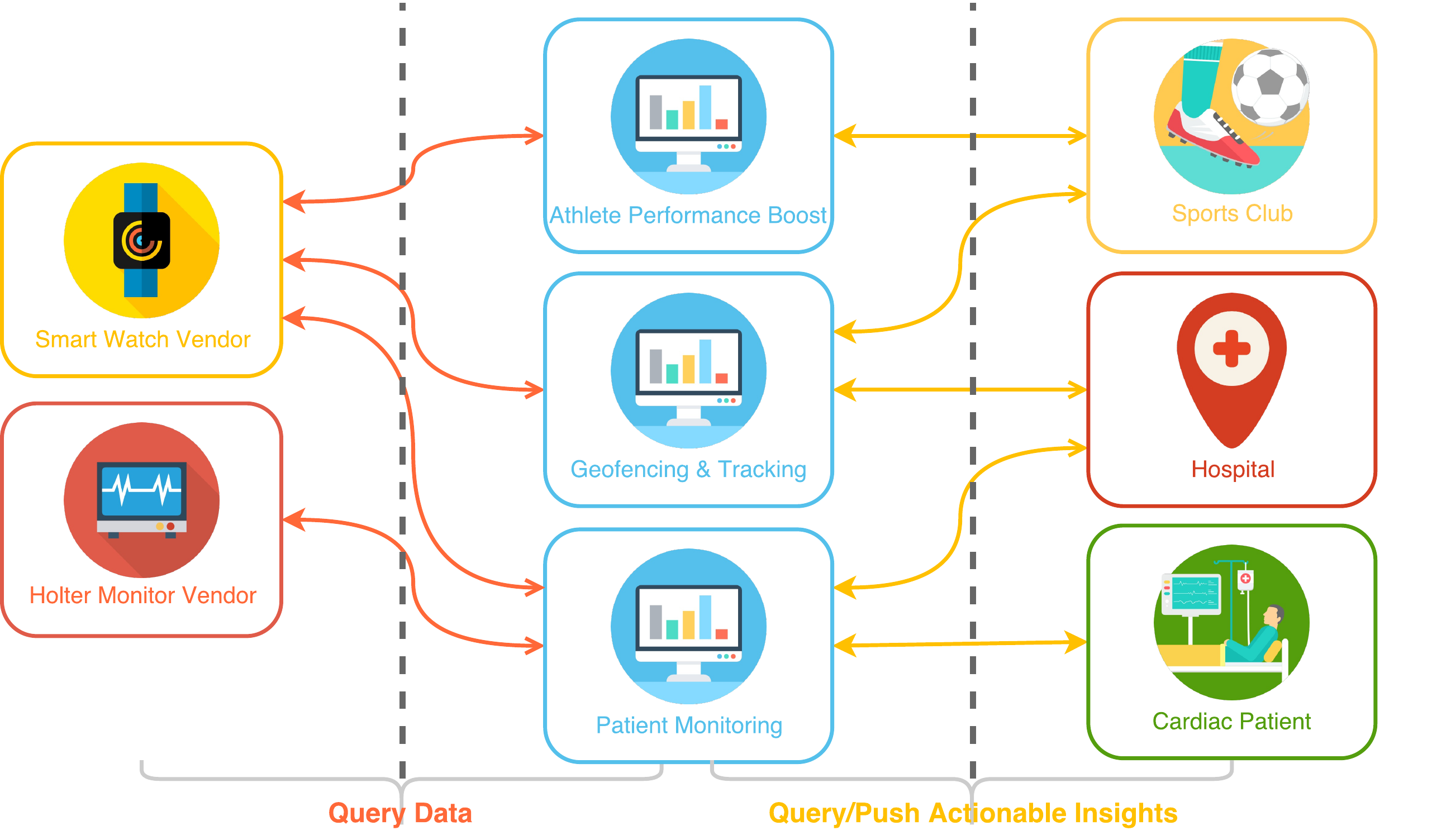}
  \caption{Example benefit network with multiple stakeholders}
  \label{fig:general}
\end{figure*}

Proposed idea facilitates an open environment with a low barrier, where businesses or regular consumers will be able to get the services or information from multiple providers. These services may be acquired in exchange for sharing device data with consent, instead of explicitly paying for the services. Proposed marketplace is not designed for time-critical systems or services that need complete user privacy or had to comply with law and data protection requirements such as General Data Protection Regulation (GDPR) (Regulation (EU) 2016/679)~\cite{eugdp}.

In short, by creating a common, decentralized and trustless infrastructure it will be possible to provide a) an always-on data store for IoT manufacturers b) a searchable data marketplace for AI/ML companies. In this paper, we aim to give insights of how such a solution can be built by using blockchain technology and lay out the mechanics and governance guidelines for such a system.

Organization of this paper is as follows: an overview of the IoT data marketplace concept and its benefits are presented in the next section (Section~\ref{sec_marketplace}). Then, the requirements and limitations for such a system is described in Section~\ref{sec_requirements}. In "Candidate Platforms" (Section~\ref{sec_candidates}), prospective blockchain platforms for implementing the data marketplace are evaluated. "Implementation Concepts" (Section~\ref{sec_concepts}) and "Smart Contract" (Section~\ref{sec_smartcontract}) sections focus on the implementation while providing insights on key features of the contract and the skeleton code, respectively. Implementation sections are followed by a discussion (Section~\ref{sec_discussion}) on the pain points and improvement opportunities regarding the proposed system.

\section{IoT Marketplace on Blockchain}
\label{sec_marketplace}
Having a decentralized IoT data platform has multi dimensional benefits for all of the contributing parties as detailed in this section. Such a system will not only provide economical benefits, but also technical and user-facing benefits as well. As seen in Figure~\ref{fig:general}, IoT data can be collected, processed and finally consumed by different parties. In this figure, there are two IoT devices: a smart watch and a holter monitor. The device manufacturer must have already listed the data of these devices by using the proposed marketplace. Three different AI/ML providers buy the data of these devices and process them with their algorithms to produce insights such as tracking a team\textquotesingle s or an athlete\textquotesingle s performance, location of a certain or a group of individuals or non-real time monitoring of patients. As a result, these insights may be exploited by individuals, companies or organizations like hospitals, sport clubs or a single patient. In the end, this approach will democratize the way data is managed and will accelerate IoT adoption, creating a positive feedback loop.

As a side note, although the concept of the decentralized, trustless data marketplace is fairly new, IOTA platform~\cite{iota} has also proposed a data marketplace. In the last part of this section, previous experiments and proposals for building such systems are reviewed.

\subsection{Benefits}

\subsubsection{Technical Benefits}
Having a common, blockchain-based data backend has clear technical benefits for all the parties in the system:
\begin{itemize}
  \item IoT manufacturers don’t have to create and maintain cloud backends for sensor data, because our solution will provide the necessary data analytics services over blockchain.
  \item IoT manufacturers will use well-tested, maintained and optimized code for their devices to interact with Swarm and Ethereum. Software development cycles will be reduced and time-to-market for new products will decrease significantly.
  \item AI/ML providers may be able to tap into a vast pool of data that they are unable to reach before. Therefore AI/ML solutions will improve due to increased amount of training and test data.
  \item Consumers of the actionable insights, e.g. businesses, organizations and end users will be able to build new kinds of products and services. They will be able to browse through a vast library of behavioural patterns that are created by AI/ML providers.
\end{itemize}

\subsubsection{Economical Benefits}
Obviously, the most direct benefit of the system will be in economic terms. Specifically, IoT manufacturers, AI/ML providers and end users will benefit directly and indirectly as detailed below:
\begin{itemize}
  \item IoT manufacturers will be able to monetize consented user data which may ignite a new wave of business models where IoT device costs may reduce to zero due to the subsidies coming from data monetization.
  \item AI/ML providers will be able to sell actionable insights to businesses and users.
  \item Businesses may finally provide predictive, prescriptive and adaptive solutions to their customers.
  \item End users may be able to use IoT devices and services freely in exchange for their consent on data usage. The scope of data collection is determined by both user’s consent and legal framework that allows or denies it.
\end{itemize}

\subsubsection{User-facing Benefits}
In general, there is a skeptical approach to systems where user data is collected and processed, mostly due to privacy concerns. However this is not a zero-sum game. A transparent and trustless data marketplace that only contains consented data may result high quality products and services for consumers.
\begin{itemize}
  \item In the proposed model, there are two factors that will reduce IoT device prices: the first one is the lack of cloud backend management burden and the second one is the data monetization capability. Assuming IoT device manufacturers reflect decreased costs to device prices, there will be a net incentive for end users to use these products and services. It is likely that free-to-use or free-if-consented business models bloom in every industry.
  \item Such a data marketplace will bring democratization by  means of data access. This is one of the problems that Web technologies have created, considering the enormous amount of personal data accumulated and controlled by Google and Facebook. In our proposed system, data is intertwined with protocol. Hence, there is no way to monopolize the data in the system.
  \item A rich ecosystem filled with IoT manufacturers and AI/ML providers will create a universal library, consisting raw and processed data. It will be open to anyone who wants to search for correlations for any set of inputs and outputs.
\end{itemize}

\subsection{Experiments}
At the end of 2017, IOTA announced that it is going to support a decentralized marketplace, "to open up the data silos that currently keep data limited to the control of a few entities. Data is one of the most imperative ingredients in the machine economy and the connected world"~\cite{iotamarket}. Although the exact number of IoT vendors and devices that use this platform is not exactly known, IOTA\textquotesingle s marketplace approach shares a lot of goals and ambitions with the proposed design. The main difference in our proposed approach is that there is no entity that oversees the marketplace in any form. Only a transparent, independent and auditable smart contract is in place, which takes care of connecting data providers and consumers. The data marketplace application is decentralized and trustless in itself.

It is also imperative to mention about a previous experiment, named "Contract Market"~\cite{contractmarket} where users can subscribe to IoT devices. Hence, IoT vendors are able to manage them via a smart contract. However, the details regarding where and how the data is stored and accessed, or how the system economics would work is missing.

There are also non-blockchain based attempts to create a common data platform. One of them is "Big IoT Marketplace"~\cite{bigiotmarketplace} by the European IoT Platforms Initiative. Big IoT Marketplace is a platform where IoT data providers will be able to sell their data. However, it is both centralized and does not provide a generic method to store and access data.

\section{System Requirements and Limitations}
\label{sec_requirements}
Finding the right blockchain platform for implementing an IoT data infrastructure requires consideration of multiple key aspects: data storage mechanism (on-chain or off-chain), tools and capabilities for creating an IoT data platform and financial incentivization for sustainability.

\subsection{Data Storage}
“Data storage mechanism” is a general term to describe how IoT sensor data is pushed and where it will be stored. First generation public blockchains have a cap on number of transactions, either in form of block size (Bitcoin) or gas limit per block (Ethereum). Pushing IoT data directly into these systems is not feasible for the majority of IoT applications, due to the high amount of transactions and the high amount of data. Bitcoin is able to process ~4.5 transactions per second (2704 transactions per block on 21th of December, 2017)~\cite{btctx} and Ethereum is able to clear out ~15.6 transactions per second (1349890 transactions on 4th of January, 2018)~\cite{ethtx} at their peak. On the other side, there are private blockchain platforms like Hyperledger that has low latency requirements for consensus but do not fully satisfy decentralization goals. Benchmarking of Hyperledger platform shows that it fails to scale beyond sixteen nodes~\cite{dinh2017blockbench}.

Quorum~\cite{quorum} and Corda~\cite{brown2016corda}, which are both permissioned and blockchain-inspired platforms targeting financial institutions, proposing a different model where data is not stored publicly on blockchain. Instead, data is kept off-chain by the participating peers (financial institutions) and the consensus function is designed to ensure agreements among interacting parties. Although this approach may be practical for financial institutions in terms of creating "business flows", it eliminates one of our design goals where IoT device manufacturers use this system as an "always-on data store". In addition, there are custom blockchain platforms targeting IoT and decentralized application development, such as IOTA~\cite{popov2016tangle} and EOS~\cite{eoswp}, which will be analyzed separately below.

Although pushing the complete IoT data into blockchain is problematic, it should be possible to push a so-called “file handle”, that is tied to a specific IoT data chunk. Hence, our proposed data marketplace targets non-real time and non-critical IoT systems that push monitoring data to the data backend in large time intervals (>30 mins). However, this approach will need a secondary decentralized file storage layer. IPFS~\cite{benet2014ipfs} and Swarm~\cite{swarm} are two prominent alternatives that can be used for this purpose. Both technologies are peer-to-peer (P2P) with decentralized file transfer systems in which files are addressed by the hash of their content. Moreover, they are compatible with the concept of edge computing if IPFS or Swarm nodes are executed on IoT gateways. On top of that, highly used data sources will be retrieved with low latency as mentioned in Swarm guide: “Nodes cache content that they pass on at retrieval, resulting in an auto scaling elastic cloud: popular (oft-accessed) content is replicated throughout the network decreasing its retrieval latency”~\cite{swarmguide}.

\subsection{Decentralized Application}
Proposed IoT data platform should be completely decentralized and always in working condition in all circumstances. Therefore not only the financial part of the trade, i.e., the transactions, but also the application logic of the platform should be in the blockchain. As a result, this narrows down the list of blockchain platforms to the ones that utilize smart contracts to ensure an always-on decentralized platform.

\subsection{Financial Incentives}
There are already clear benefits for IoT device manufacturers and AI/ML providers to use the proposed system. IoT device manufacturers will be able to break free of developing and maintaining a cloud backend. Besides, they will be able to sell collected data in an open marketplace. AI/ML providers, on the other hand, will be able to access a vast data library where they can browse and buy as much as they can afford. In addition, nodes in the decentralized storage network should also be incentivized in order to keep bulk IoT sensor data available, at least based on their usage~\cite{swarm}. This mechanism is similar to Amazon Web Services (AWS) Simple Storage Service (S3) in terms of functionality. Yet, it consists of multiple independent peers committing their resources instead of a single entity, where they are rewarded based on their contribution. Storage incentivization can be done only if decentralized storage system is deeply integrated with the blockchain client. Having a built-in currency is a vital tool for embedding incentives at the transaction level. Unfortunately, permissioned blockchains like Hyperledger and Corda lack this mechanism.

\section{Candidate Platforms}
\label{sec_candidates}
The requirements and the challenges for such a system is detailed in the previous section. Now we will study two available candidate platforms: one is customized for addressing IoT needs (IOTA) and the other one is proposed for decentralized application development (EOS). Then, we will briefly describe our proposal, Ethereum and Swarm platform.

\subsection{IOTA}
IOTA is a relatively new project which uses "Tangle", a directed acyclic graph data structure to store transactions. It aims to provide a decentralized infrastructure and a data marketplace for IoT devices~\cite{popov2016tangle}. However due to centralization concerns and persistent storage needs (permanode), IOTA is not picked as the implementation platform for the time being. 

"Coo" (Coordinator), which is a full node controlled by IOTA Foundation, is employed to clear out transactions. If "Coo" is down by any reason, IOTA network stops working. IOTA plans to shutdown "Coo" when the system is able to resist to a \%34 attack. However, at the time of this writing, it is still on. In addition, IOTA uses a mechanism called "snapshot" where they prune history of transactions and the attached data in order to prevent bloat. As a result, IOTA full nodes will not be storing any data by default (even a pointer to an external file) except the account balance. In order to access persistent data, IoT vendors should run so-called “permanodes” that store all the data starting from the genesis block. This will be a huge burden for IoT vendors in terms of storage compared to just incentivized independent Swarm nodes for storage in a Ethereum-Swarm setup. As a part of the announced milestones, IOTA is planned to take automatic snapshots.

\subsection{EOS and EOS Storage}
EOS is another blockchain project aiming to create scalable, decentralized applications on top of an existing blockchain architecture~\cite{eoswp}. EOS project addresses important aspects like creating peer-to-peer terms of service agreements, separating authentication from application. These aspects are very important if the aim is to create a decentralized peer-to-peer data marketplace. Similar to the authors’ line of thinking, EOS Whitepaper~\cite{eoswp} emphasizes that the piece of data to be stored in blockchain should be relevant to the application. In other words, instead of the content itself, i.e. bulk IoT data, a pointer to it should be stored in blockchain. Just like Swarm, EOS recently proposed a decentralized storage layer built on top IPFS technology~\cite{eosstorage}. In order to have a replicated file on EOS Storage, two transactions should be processed, one for creating the file on the blockchain and the other one for the confirmation of a successful upload~\cite{eosstorage}. In general, EOS is a well-thought platform tailored for the needs of the next generation of application developers, However, for our specific use case EOS storage mechanics will double the amount transactions needed to store the file handles.

\subsection{Ethereum and Swarm}
Ethereum, being the first decentralized application platform, has already established itself a high ranking among cryptocurrencies and sparked developer interest with its decentralized application platform. Ethereum currently offers a widely used programming language, called “Solidity”, and a complete web based development environment, called “Remix IDE”. In addition, Ethereum is deeply integrated with Swarm, a decentralized, torrent-like storage service. As the result of this deep integration, Swarm nodes can be financially incentivized directly from Ethereum. Lastly, Ethereum platform’s currency “Ether” is widely used which makes instant trades to fiat currency possible. Such a feature may accelerate adoption of the proposed system exponentially.

Based on the aspects detailed above, Ethereum is selected as the blockchain and Swarm as the decentralized storage platform based on the maturity of the platforms and deep integration with each other.

\section{Implementation Concepts}
\label{sec_concepts}
This section will first go over the concepts used in the development of our smart contract. Designing a generic IoT backend on the blockchain requires some challenges which can be listed as follows:
\begin{itemize}
  \item a flexible querying mechanism for data consumers (filter data by vendor, sensor type, geo-location, time)
  \item a voting system to rank data sources
  \item a token-based economy where marketplace payments are not exposed to heavy market fluctuations
  \item payment channels to execute instant transfers
\end{itemize}

\subsection{Data-as-a-Contract}
Implementing the IoT data marketplace as a smart contract, i.e. a decentralized application deployed as a part of blockchain, facilitates a transparent data collection and sharing environment. In addition, by being trustless, blockchain infrastructure inherently provides a global safe-trade environment. Quoting Nick Szabo, “trusted third parties are security holes”~\cite{nickszabo}.

The presence of such a system proactively eliminates operational risks of IoT device manufacturers, as there will be no need to develop and maintain an actual data backend. In our previous research, we demonstrated how the blockchain platform can be used for integrating IoT devices~\cite{Ozyilmaz:2017:ILI:3125503.3125628} and creating a generic data backend~\cite{cesoc}. IoT device manufacturers may maintain a custom data backend for certain purposes like privacy, but in any case, a blockchain-based system will enforce transparency and data democratization.

Nowadays, it is very common for cloud provides to create solutions labeled as Software-as-a-Service (SaaS) or Platform-as-a-Service (PaaS) that offer their customers pay-per-use access. Following that model, consumers of the IoT data such as AI/ML providers, or consumers of the actionable insight such as business, organizations, regular end users who give consent to data sharing, will pay as much as they use the provided services.

\subsection{Geographical Data on Blockchain}
One of the key features for Internet of Things is geolocation where the interaction with environment takes place. There are many researches for determining geolocation of IoT end devices without trusting the location information from IoT device or IP packets, which may be blocked or compromised by an adversary~\cite{islam2017determining}. Still, detecting the geolocation of an IoT device is not ethical due to violation of privacy. Actually, it is possible to verify the geolocation of an IoT end-device by the consensus of nearby IoT end devices on a mature ecosystem unless more than 33\% of devices behave maliciously.

Hence, we propose to use GeoHex that divides whole world map into hexagons and map these hexagons with strings~\cite{geohex}. It is very practical when it comes to searching for nearby geolocations. Whereas most of geolocation systems require floating point arithmetics, GeoHex limits the geolocation to a string of at most 17-bytes. Considering its importance for a data scientist in picking-up data from the market, we prefer using GeoHex due to its ease of use in querying geolocations.

\subsection{Validation and Feedback}
Voting is a straightforward feedback mechanism that is used by the AI/ML providers to rank the quality of the IoT device manufacturers. Consumers of the data marketplace will mark bad providers, which will in turn increase the overall data quality in the system. Voting is one of the early concepts explored in blockchain systems and due to its immutable and trustless nature, such applications proved to be working successfully.

\subsection{Data Tokens}
Ethereum tokens are ERC20-compatible smart contracts that can act like a currency on top of Ethereum~\cite{erc20}. By creating Ethereum tokens, it is possible to define a custom currency, which can be used to interact with the proposed, underlying system. In short, smart contracts can be extended to define their own economic model. A direct benefit of such an abstraction is the isolation of the token value from the price fluctuations of Ethereum.

The proposed data marketplace uses Ether (Ethereum\textquotesingle s currency) as the medium of exchange. However, by extending the smart contract to support ERC20 standard, proposed marketplace will be able to offer a custom token to be used as a currency, therefore providing a stable and deterministic data pricing. Even though it is purely economic, these type of changes are required to facilitate mass adoption of the proposed system.

\subsection{Payment Channels}
Blockchain systems process transactions by packaging them into blocks, which inherently adds latency expressed in block creation time. Average block creation time is 14 seconds in Ethereum. This latency, however, is not ideal for scenarios where high amount of small payments are taking place between two parties. In order to address this issue, an off-chain scaling solution called \textit{payment channels} has emerged. Payment channels are near-instant and low-fee payment networks, complementing the original blockchain platform. Currently, Lightning Network~\cite{poon2016bitcoin} provides this service for Bitcoin and Bitcoin-variant currencies, and Raiden Network~\cite{raiden} provides it for Ethereum blockchain and works with any ERC20 compatible token.

The proposed data marketplace will benefit from payment channels, because marketplace should offer instant exchange of token and data with its dedicated ERC20 token. By using Raiden micropayment network and instant token transfers, it will become possible to introduce pay-as-you-go or subscription-based solutions for data consumers.

\begin{figure*}[t!]
  \centering
  \includegraphics[scale=0.8]{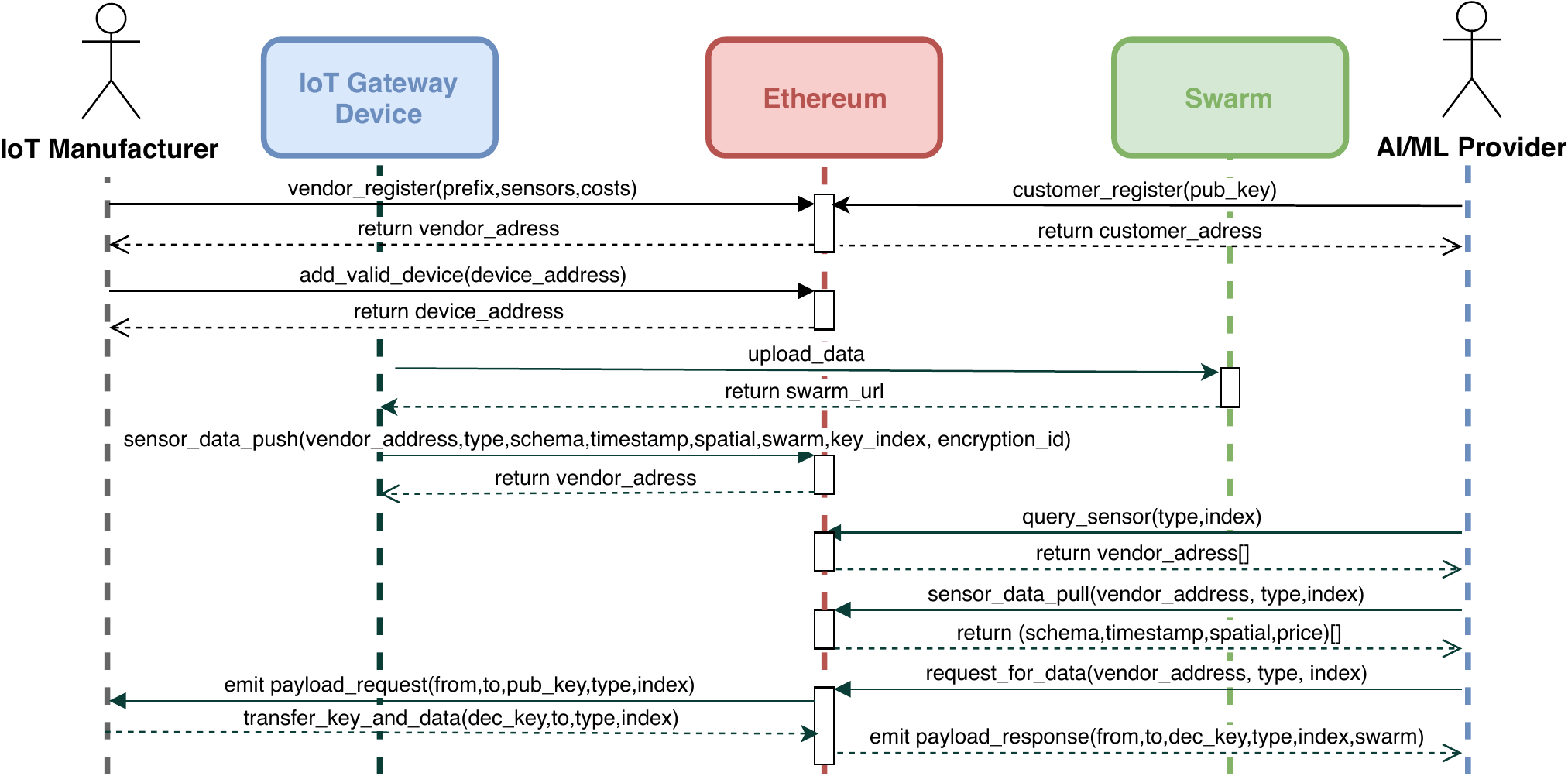}
  \caption{Data Flow Sequence Diagram}
  \label{fig:sequence}
\end{figure*}

\begin{listing*}
\begin{minted}[mathescape, linenos, numbersep=5pt, frame=lines, framesep=2mm, fontsize=\scriptsize]{csharp}
    function vendor_register (string prefix, uint[] sensors, uint[] costs) public returns (address);
    function customer_register (string pub_key) public returns (address);
    function add_valid_device (address device_address) public returns (address);
    function vendor_length () public view returns (uint length);
    function get_vendor (address addr) public view returns (string prefix);
    function vote_for_vendor (address vendor_address,uint vote) public returns (uint);
    function query_sensor (uint sensor_type, uint index) public view returns (address result);
    function sensor_data_push (address vendor_address, uint sensor_type, string schema, uint timestamp,
				string spatial, string swarm, uint key_index, uint enc_id ) public returns (address);
    function sensor_data_pull (address vendor_address, uint sensor_type, uint index)
				public view returns (string schema, uint timestamp, string spatial, uint price);
    function sensor_data_length (address vendor_address, uint sensor_type) public view returns (uint len);
    function get_sensor_price (uint sensor_type_index) public view returns (uint);
    function update_sensor_price (uint sensor_type, uint price) public returns (uint);
    function request_for_data (address vendor_address, uint sensor_type, uint index) public returns (address);
    function transfer_key_and_data (string dec_key,address _to, uint sensor_type, uint index) public returns (string);
\end{minted}
\caption{Data Marketplace Core Functions}
\label{list:funcs}
\end{listing*}

\section{Smart Contract}
\label{sec_smartcontract}

\subsection{Overview of Development Environment}
Ethereum blockchain with a built-in Turing-complete programming language allows us to write smart contracts~\cite{ethereumwp}. In this paper, the smart contract implementation is done in Solidity which is designed to target Ethereum Virtual Machine (EVM). During the smart contract development process, we used web-based Remix IDE, which contains Solidity compiler and debugger. Ethereum client version geth 1.7.3, Solidity version 0.4.19 and Remix IDE Online version 0.1.3 are used for development.

The contract written in Solidity generates two components: the bytecode to run on EVM and the Application Binary Interface (ABI). Bytecode runs whenever a function is called from the application, and stored into Ethereum blockchain under contract address. ABI defines the structures and functions that can be invoked explicitly. In other words, ABI grants access to call functions in smart contracts. To sum up, three requirements should be satisfied to interact with a smart contract: 1) Bytecode must be deployed to blockchain 2) Address of bytecode must be known 3) ABI of smart contract must be known.

Our proposed system will be able to provide IoT device data to many users when the system reaches its maturity. The relation between the data and its users is similar to one-to-many relationship in traditional databases. Based on this relationship, our platform can be seen as a decentralized exchange platform that requires more reading operation from smart contract than writing to it.

\begin{listing}
\begin{minted}[mathescape, linenos, numbersep=5pt, frame=lines, framesep=2mm, fontsize=\scriptsize]{csharp}
    /* payload from a specific sensor type */
    struct payload {
        address device_id;
        uint timestamp;
        string swarm;
        string schema;
        string spatial;
        uint key_index;
        EncryptionScheme encryption_scheme;
        string encrypted_key;
    }
    /* everything about vendors */
    struct vendor {
        string prefix;
        // vendor supported sensor types
        mapping(uint => bool) types;
        // unit prices for every sensor type
        mapping(uint => uint) prices;
        // payload from a specific sensor type
        mapping(uint => payload[]) payloads;
        // devices belong to specific vendor
        mapping(address => bool) devices;
        // total count of votes
        uint votes;
    }
    struct customer{
        payload[] paid_arr;
        mapping(address => bool) vote_map_used;
        string pub_key;
    }
    mapping(address => vendor) private vendor_map;
    mapping(address => customer) private customer_map;
    address[] private vendor_arr;
    mapping(address => uint) balances;
\end{minted}
\caption{Data Marketplace Core Data Structure}
\label{list:code}
\end{listing}

\subsection{Action Flow}

The main stakeholders of the application are “vendors” and “customers”, who correspond to an IoT Manufacturer and an AI/ML provider, respectively, in Figure~\ref{fig:sequence}. When a vendor wants to get economic benefits from devices, it creates a new registry on the application by calling \textit{vendor\_register} (Listing~\ref{list:funcs}, line 1). For blocking an unauthorized device pushing data to market on behalf of the vendor, vendor must declare its device addresses by using \textit{add\_valid\_device} method (Listing~\ref{list:funcs}, line 3).Then, any valid (registered by vendor) device can push data to the system by stating the vendor, pre-defined schema, file handle, timestamp, and geolocation (Listing~\ref{list:funcs}, line 8).

In this manner, devices can upload many datasets from different sensor types into the system. Then any user, such as an AI/ML provider, can query data sets of a sensor by calling \textit{query\_sensor} (Listing~\ref{list:funcs}, line 7). It returns list of vendors who own the datasets of the queried sensor type. From this point on, the user selects a vendor and the application calls \textit{sensor\_data\_pull} to have more descriptive details such as  timestamp, geolocation or schema of the sensor data (Listing~\ref{list:funcs}, line 10). After desired dataset is matched, payload data can be claimed by calling \textit{request\_for\_data} (Listing~\ref{list:funcs}, line 15). When the user retrieves the data, voting option for the user is enabled to evaluate the vendor. Through \textit{vote\_for\_vendor} (Listing~\ref{list:funcs}, line 6), the user is able to vote as up or down according to his/her experience.

\subsection{Data Structures and Optimization}

IoT device data is a form of digital asset which is controlled by the smart contract in data market application. It is considered as a digital asset because data collection is a costly operation for IoT vendors and the collected data provides value for businesses. IoT device data or namely the \textit{payload} (Listing~\ref{list:code}, line 2-11) is the fundamental structure, around which the whole ecosystem gets shaped. Storing this data directly on blockchain creates lots of transactions and incurs high financial costs. Instead, IoT device data is uploaded from gateway to Swarm file system in an encrypted form. Swarm client returns file handle, which is cryptographic hash of the data.  The file handle is unique identifier and address of data. Data schema, which will be used for parsing the payload, is an important concern for AI/ML providers. Therefore, it is also included in the \textit{payload} structure. In addition, the identifier of the device that uploads the data, the name of the encryption scheme and the index of the encrpytion key is also present in the \textit{payload} structure. Details on encryption and security will be given in the next subsection "Encrpytion and Data Security"~\ref{sec_encryption}.

Second structure in the contract is \textit{vendor} (Listing~\ref{list:code}, line 13-25). Vendor is located at the lowest level in the stack-like approach which is shown in (Figure~\ref{fig:stack}). To show human-readable names instead of addresses, we store \textit{prefix} for each vendor. In our implementation, we used a unique number instead of a string to represent each sensor type. For example, “1” for smart watch data, “2” for holter monitoring data, etc. By doing so, we decrease the cost of transactions~\cite{bytearray}. Based on this representation, we used sensor type as key for \textit{types}, \textit{prices}, and \textit{payloads} mappings. \textit{Types} store for which sensor types registered by the vendor. \textit{Prices} store the corresponding prices for each sensor type. A vendor has the ability to push multiple payloads per sensor type, with each \textit{payloads} store array of payload structures (Listing~\ref{list:code}, line 20). Device addresses that are allowed to push data in the name of the vendor are also stored for automatization. This allows the IoT device to export its data to Swarm with a script and add a new payload to the application with a file handle on behalf of vendor. Reliability and convenience of datasets can be provided with a voting mechanism, so \textit{votes} is also defined in the vendor structure as a field.

The last structure of the system is \textit{customer}, which can be an AI/ML provider or an individual user. It stores the public key of the AI/ML provider to be used in later stages during data decryption. For browsing and voting purposes, \textit{paid\_arr} and \textit{vote\_map\_used} mappings are defined in this structure (Listing~\ref{list:code}, line 26-30). \textit{Vote\_map\_used} stores address of vendor as the key and a boolean value that shows whether the customer has the right to vote or not.

While implementing functions, our main concern was to minimize gas cost which is spent on each execution of opcodes in EVM. Considering this cost, we avoided loops and mapped data structures accordingly. We used two mapping structures as global variables (Listing~\ref{list:code}, line 31-32) for getting or setting any field within stakeholders (customers and vendors) of the system. Instead of storing whole vendor structs in an array, we stored addresses of them for querying sensors and corresponding prices (Listing~\ref{list:code}, line 33).

Any operation like registering as a vendor with a fake ID or adding random devices to vendor\textquotesingle s space is punished by Ethereum network itself. Therefore, we did not implement an additional blocking mechanism.

\subsection{Encryption and Data Security}
\label{sec_encryption}
Swarm file handles are openly visible on the blockchain. In order to prevent a non-paying user get all the Swarm handles and fetch the corresponding files, IoT data uploaded to Swarm should be encrypted with a symmetric key before the upload. To ensure that, payload metadata should contain the name of the encryption scheme (DES, AES) and a key index beside the device identifier.

We propose that IoT device vendors store the master keys of their gateway devices and configure the devices to create new symmetric encrpytion keys for each upload using a hierarchically deterministic method as it is done in BIP32 Wallets~\cite{bip32}. This way, every Swarm file will be encrypted with a different symmetric key and IoT device vendor will be able to calculate any given key by using the master key and the key index provided inside the metadata. IoT device manufacturers store and manage master keys per device even today, especially for low-power, long range protocols like LoRa where a master key is used to encrypt data messages on the field ~\cite{sornin2015lorawan}.

If a data consumer wants to buy a certain chunk of IoT data, the payment and acquisition of the data will happen as described in previous sections. The decryption process of the acquired data will happen as follows, where the whole process can be automated by Javascript code interacting with the Ethereum client:
\begin{enumerate}
\item AI/ML provider will request data and pay for it by using a smart contract function
\item IoT device vendor will be notified by the event invoked by that call, passing the address of the AI/ML provider
\item IoT device vendor will use the address to get the public key of the AI/ML provider
\item IoT device vendor will calculate the symmetric key that is used to encrypt that particular Swarm file by using that device\textquotesingle s master key and key index
\item IoT device vendor will encrypt symmetric key with AI/ML providers public key and create a transaction
\item AI/ML provider will receive the encrypted symmetric key, decrypt it using its private key and then decrpyt the Swarm file using the symmetric key
\end{enumerate}

\section{Discussion}
\label{sec_discussion}
\textbf{Encryption:} Ethereum blockchain does not store data in encrypted form. Similarly, the proposed data marketplace does not impose any restrictions on the IoT data uploaded to Swarm. It only sets the mechanics between data vendors and data consumers. In the current design, it is assumed that the IoT data sent to the platform is anonymized due to the rules and regulations, like GDPR, that IoT device vendors are facing. Although IoT device manufacturers may decide to encrypt data and share the keys with data consumers by using an off-chain method in the current system, it is not very practical. It is planned to extend the data marketplace to support some form of encryption where encrypt/decrypt operations can be conducted in a decentralized manner.

\textbf{Real-Time Systems:} Public blockchain systems add blocks, i.e. packaged transactions, to the blockchain at every block creation interval on average. On top of this, there is a block propagation delay which adds additional latencies if a data consumer tries to follow a real-time data feed using the data marketplace. Therefore, with the current consensus functions on widely used public blockchain platforms, proposed solution does not support real-time or safety-critical applications due to high latencies.

\textbf{Data Collection and Consent:} European Union has data protection requirements such as General Data Protection Regulation (GDPR) (Regulation (EU) 2016/679)~\cite{eugdp} already in place, so IoT device manufacturers should comply with current rules and regulations as data providers. Although data marketplace smart contract does not store any user data (just Swarm handles), data replication on Swarm filesystem should be managed by device vendors.

\section{Conclusion}
In our previous research, we have already explored ways to integrate low-power IoT devices to a blockchain-based infrastructure~\cite{Ozyilmaz:2017:ILI:3125503.3125628} and created a decentralized data backend~\cite{cesoc}. In this paper, we extend that goal to a broader data marketplace involving multiple parties, targeting non-real time, non-critical IoT applications. Creating a decentralized and trustless platform for storing and accessing IoT data will positively impact IoT device manufacturers, AI/ML providers, and, obviously, the end-users. Such a marketplace will democratize access to consented data and increase both service quality and variety of offerings, which will turn out to be beneficial for the users in the end.

A proof-of-concept data marketplace is implemented as a smart contract on Ethereum platform and uses Swarm as its storage system. It provides a flexible querying mechanism for data consumers and contains a voting mechanism for eliminating unreliable data providers. Smart contract code is open sourced on GitHub as "IDMoB: IoT Data Marketplace on Blockchain"~\cite{idmob}.

\section*{Acknowledgment}
This project has been partially supported by Scientific Research Fund of Bogazici University under grant number: 13500

\bibliographystyle{IEEEtran}
\bibliography{IEEEabrv,cv-iot}

\end{document}